\documentclass[12pt]{article}

\usepackage[utf8]{inputenc}
\usepackage{charter}
\usepackage{fullpage}
\usepackage{hyperref}
\usepackage{graphicx}
\usepackage{amssymb,amsmath}
\usepackage{lipsum}
\usepackage[sort&compress,numbers]{natbib}
\usepackage{hyperref}
\hypersetup{hidelinks}
\usepackage{wrapfig}
\usepackage[usenames,dvipsnames,table]{xcolor}
\usepackage[total={6.5in,9in},top=1in,headsep=0.1in,headheight=1in,]{geometry}
\usepackage{lineno}
\usepackage{ads}
\usepackage{siunitx}
\usepackage{geometry}
\usepackage{fullpage}
\usepackage[T1]{fontenc}    
\usepackage{bm}	
\usepackage[normalem]{ulem}
\usepackage{caption}
\usepackage{subcaption}
\usepackage{verbatim}
\usepackage{lineno}
\usepackage{url}    
\usepackage{booktabs}       
\usepackage{amsfonts}       
\usepackage{nicefrac}
\usepackage{natbib}
\usepackage{xcolor}

\def\Neff{N_{\rm eff}}
\mathchardef\mhyphen="2D

\newcommand\snowmass{
\begin{center}
  \rule[-0.2in]{\hsize}{0.01in}\\
  \rule{\hsize}{0.01in}\\
  \vskip 0.1in
  Submitted to the Proceedings of the US Community Study\\ 
  on the Future of Particle Physics (Snowmass 2021)\\
  \rule{\hsize}{0.01in}\\
  \rule[+0.2in]{\hsize}{0.01in}\\[-2em]
\end{center}
}
 
 \usepackage[firstpage=true]{background}
\backgroundsetup{contents={\parbox{6.5in}{\snowmass}}, scale=1,placement=top,opacity=1,color=black,position={3.25in,1.2in}}

\usepackage{fancyhdr}
\fancypagestyle{plain}{%
  \fancyhf{}%
  \fancyhead[C]{}
  \fancyfoot[C]{\thepage}
}

\fancypagestyle{empty}{%
  \fancyhf{}%
  \fancyhead[C]{{\it Dark Matter Physics from the CMB-S4 Experiment}}
  \fancyfoot[C]{\thepage}
}

\usepackage{authblk}

\author[1]{Cora Dvorkin\footnote{cdvorkin@g.harvard.edu}}
\affil[1]{Department of Physics, Harvard University, 17 Oxford Street, Cambridge, MA 02138, USA}
\author[2,3]{Ren\'{e}e Hlozek}
\affil[2]{Dunlap Institute for Astronomy and Astrophysics, University of Toronto, 50 St. George Street, Toronto, ON M5S 3H4, Canada}
\affil[3]{Department of Astronomy and Astrophysics, University of Toronto, 50 St. George Street, Toronto, ON M5S 3H4, Canada}
\author[4]{Rui An}
\affil[4]{Department of Physics and Astronomy, University of Southern California, Los Angeles, CA, 90089, USA}
\author[5]{Kimberly K.~Boddy}
\affil[5]{Department of Physics, The University of Texas at Austin, Austin, TX 78712, USA}
\author[6]{Francis-Yan Cyr-Racine}
\affil[6]{Department of Physics and Astronomy, University of New Mexico, Albuquerque, NM 87106, USA}

\author[7]{Gerrit S. Farren}
\affil[7]{Department of Applied Mathematics and Theoretical Physics, University of Cambridge, Cambridge CB3 0WA, United Kingdom}
\author[4]{Vera Gluscevic}
\author[8]{Daniel Grin}
\affil[8]{Department of Physics and Astronomy, Haverford College, 370 Lancaster Ave, Haverford, PA 19041, United States}
\author[9]{David J. E. Marsh}
\affil[9]{Department of Physics, King's College, London WC2R 2LS, United Kingdom}
\author[10]{Joel Meyers}
\affil[10]{Department of Physics, Southern Methodist University, Dallas, TX 75275, USA}
\author[2]{Keir K. Rogers}
\author[11]{Katelin Schutz}
\affil[11]{Department of Physics, Ernest Rutherford Physics Building, 3600 Rue University, Montréal, QC H3A 2T8}
\author[12]{Weishuang Linda Xu}
\affil[12]{Berkeley Center for Theoretical Physics, South Hall Rd, Berkeley, CA 94720, United States}

\date{}
\title{Dark Matter Physics from the CMB-S4 Experiment}

\begin{document}

\maketitle

\textbf{Endorsers:} Kevork N. Abazajian, Peter Adshead, Mustafa Amin, Denis Barkats, Darcy Barron, Amy N. Bender, Bradford A. Benson,  Colin Bischoff, Julian Borrill, Thejs Brinckmann, John E. Carlstrom, Clarence Chang, Jacques Delabrouille, Eleonora Di Valentino, Thomas Essinger-Hileman, Simone Ferraro, Jon E. Gudmundsson, Selim C Hotinli, Kevin M. Huffenberger, Bradley R. Johnson, Louis Legrand, Marilena Loverde, Yin-Zhe Ma, Tony Mroczkowski,  Moritz M$\ddot{\rm u}$nchmeyer, Michael D. Niemack, Giorgio Orlando, Alexandra Rahlin, Christian L. Reichardt, Mathieu Remazeilles, John E. Ruhl, Emmanuel Schaan, Radek Stompor, Cynthia Trendafilova, Matthieu Tristram, Gensheng Wang, Scott Watson, Edward J. Wollack, W.L. Kimmy Wu, Zhilei Xu, Andrea Zonca

\begin{abstract}
The nature of dark matter is one of the major puzzles of fundamental physics, integral to the understanding of our universe across almost every epoch. 
The search for dark matter takes place at different energy scales, and use data ranging from particle colliders to astrophysical surveys.
We focus here on CMB-S4, a future ground-based Cosmic Microwave Background (CMB) experiment, which is expected to provide exquisite measurements of the CMB temperature and polarization anisotropies. These measurements (on their own and in combination with other surveys) will allow for new means to shed light on the nature of dark matter. 
\end{abstract}

\newpage
\section{Introduction}

Cosmological and astrophysical observations have shown us that approximately 84\% of all matter in the universe is some form of non-baryonic dark matter (DM).  The nature of dark matter is still unknown, though a combination of laboratory experiments and astronomical observations have placed broad limits on the physical properties of dark matter candidates. We expect to see a huge improvement in the sensitivity of astrophysical measurements in the next decade, which will provide very deep insights for a wide range of dark matter scenarios. Of particular interest are observations of the Cosmic Microwave Background (CMB), primordial light from a time when the universe was roughly \num{380000} years old. The CMB encodes information about the earliest moments of the history of the universe and its evolution over cosmic times.  The statistics of the anisotropies in the temperature and polarization of the CMB are sensitive to the densities of the various constituents of the universe and their interactions, the cosmic expansion history, and the formation of cosmic structure. This white paper focuses on how observations with CMB-S4 will provide further insights into the nature of dark matter, enabling the possibility to distinguish among several proposed models of dark matter and the `standard' cold dark matter (CDM) picture. 

The observational future of CMB science is bright. Current state-of-the-art telescopes and detectors are being replaced with ever more sensitive instruments. The next-generation CMB-S4~\cite{cmbs4} experiment will measure the CMB in multiple frequencies to noise levels of $\approx 1~\mu\mathrm{K \mhyphen arcmin}$ with a resolution of $< 1.5~\mathrm{arcmin}$ over 70\% of the sky (in addition to measuring about 3\% of the sky to a greater depth of $\sim0.5~\mu\mathrm{K \mhyphen arcmin}$ with both $\sim25~\mathrm{arcmin}$ and $\sim2.5~\mathrm{arcmin}$ resolution). CMB-S4 will be able to probe departures from the Standard Model (SM) of particle physics through deep, high-resolution maps of the sky in CMB temperature and polarization fluctuations.  

These measurements will provide a snapshot of the universe as it was around the time of recombination, and they will also reveal the imprints of structure growth at much later times.  Gravitational lensing of the CMB~\cite{Lewis:2006fu} leads to characteristic distortions of the primary CMB maps, allowing us to statistically reconstruct maps of the integrated line-of-sight density.  Scattering of CMB photons in galaxy clusters (the Sunyaev-Zel'dovich effect)~\cite{Sunyaev:1980nv,Sunyaev:1980vz} allows for the identification of the most massive bound structures in the universe out to very high redshifts.

\begin{figure}
    \centering
    \includegraphics[width=\columnwidth]{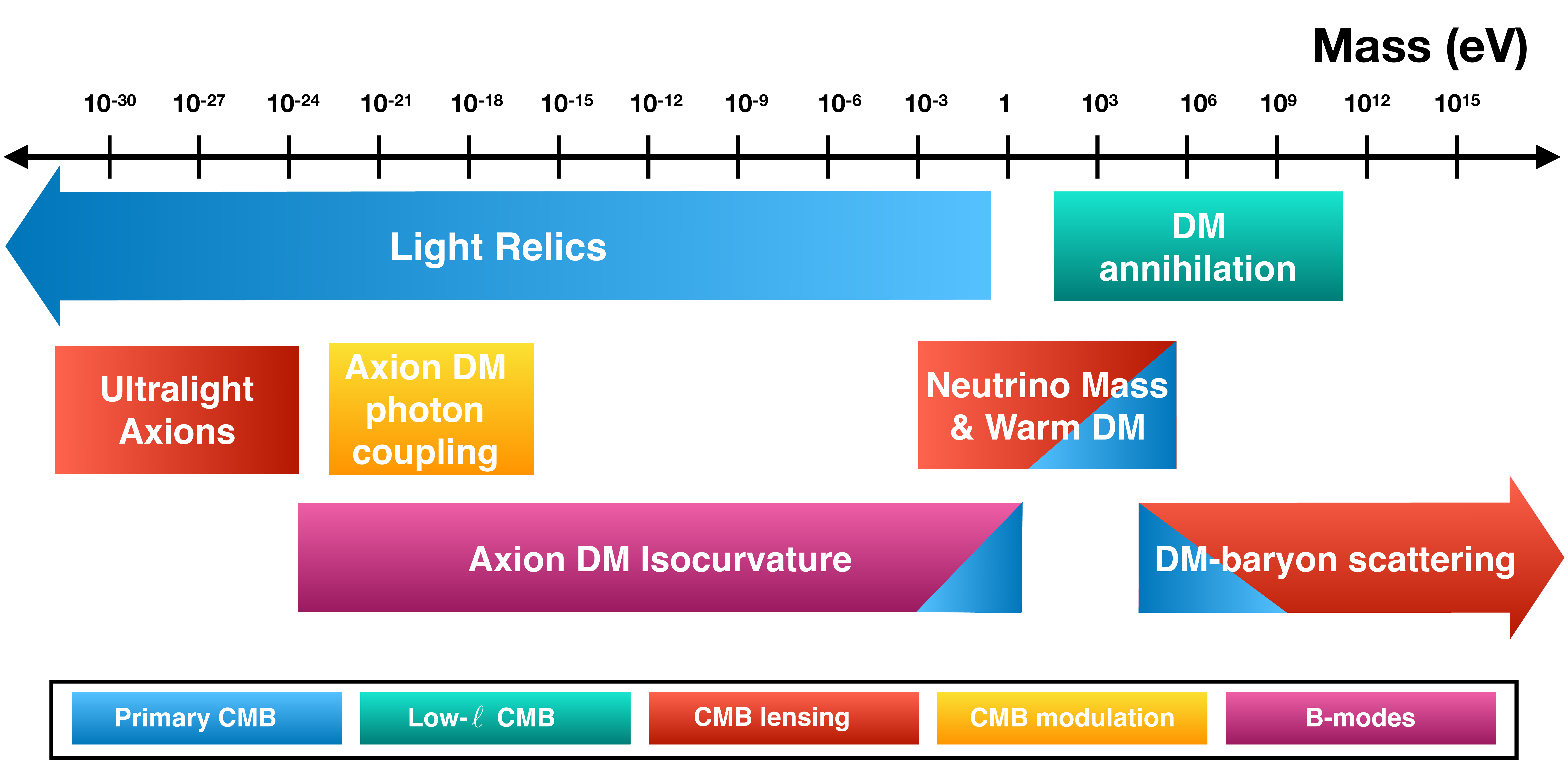}
    \caption{CMB-S4 will be able to probe light relics, axions, warm dark matter, and different dark matter scenarios. The relevant mass to which the CMB is sensitive to is shown for each case. The observable that drives the constraint is shown in different colors. Figure from Ref.~\cite{cmbs4}.}
    \label{fig:DM_mass_plot}
\end{figure}

Cosmological measurements in general and CMB measurements in particular provide insights into dark matter physics that are complementary to direct, indirect, and collider searches for dark matter.  Cosmological observables are impacted by the influence of dark matter on the entire cosmic history.  Dark matter constraints derived from cosmology do not rely on assumptions about the dark matter density in the neighborhood of the Earth or of any astrophysical object.  Furthermore, CMB observations are sensitive to regions of parameter space that out of reach of current direct searches.  Figure~\ref{fig:DM_mass_plot} provides an overview of the various means by which CMB-S4 will search for the effects of dark matter physics across a wide range of mass scales.

In the sections below, we contextualize a range of dark matter scenarios and describe the expected sensitivity of CMB-S4 to a variety of DM candidates. 

\section{Dark matter annihilation}

Weakly interacting massive particles (WIMPs) are some of the most compelling candidates for DM, where the amplitude of the self-annihilation cross section can give rise to the observed DM density today. 
When dark matter annihilates, there is an injection of energy that heats and ionizes  the baryonic gas, and this energy is inversely proportional to the DM mass. If this happens during and after recombination, it will affect the CMB temperature and polarization fluctuations \cite{Chen:2003gz,Padmanabhan:2005es}. A higher ionization fraction will increase the optical depth of the photons, and therefore it will produce a suppression in the CMB acoustic peaks. Furthermore, the extra scattering of photons will affect the polarization fluctuations at large scales. Measurements by the {\it Planck} satellite at degree angular scales provided strong bounds on annihilation and decay of DM at the sub-GeV scale \cite{Planck:2018vyg,Slatyer:2016qyl}, complementing direct detection searches that probe heavier masses. Future measurements from next-generation ground-based CMB experiments are expected to improve the current sensitivity by a factor of 2-3 \cite{Madhavacheril:2013cna,Wu:2014hta,Green:2018pmd}.  

\section{Sensitivity to thermal-relic dark matter mass}
\label{sec:themalDM}

  Current null results from direct detection searches for heavy WIMPs have inspired WIMP-like and other models for light thermal--relic DM, with sub-GeV masses \cite{2008PhRvD..77h7302H, 2008PhRvL.101w1301F, 2007PhRvD..76j3515H, 2004NuPhB.683..219B,2004PhRvL..92j1301B,2004PhRvL..93p1302H,2004PhRvD..69j1302B}. If the light thermal--relic DM particles are in thermal contact with the rest of the plasma during Big Bang Nucleosynthesis (BBN), which is the case for majority of the popular WIMP models in the current literature \cite{1996PhR...267..195J, 2012JCAP...12..027H, 2012AnP...524..479B, 2013JCAP...08..041B, 2013PhRvD..87j3517S, 2014PhRvD..89h3508N, 2015PhRvD..91h3505N, 2016PhRvD..94j3525W, 2019JCAP...02..007E, 2019JCAP...04..029D, 2020JCAP...01..004S}, and if their masses are around $\sim$0.01--20 MeV, they become non-relativistic right around the time of BBN. The resulting DM annihilation into Standard Model particles can affect the early expansion history, altering the time at which proton--to--neutron conversion, and various other nuclear processes happen, and therefore impact the production of light elements \cite{1986PhRvD..34.2197K, 2004PhRvD..70d3526S, 2004JPhG...30..279B}, including the primordial abundance of helium--4, $Y_\mathrm{p}$. Annihilation products can additionally alter the radiation content in the universe, changing the effective number of light species $N_\mathrm{eff}$ \cite{2013PhRvD..87j3517S}.
Both of these effects are imprinted on the CMB anisotropies.

The current CMB constraints on the mass of light DM were derived using \textit{Planck} data \cite{2013JCAP...08..041B, 2014MmSAI..85..175S, 2014PhRvD..89h3508N, 2015PhRvD..91h3505N, 2019JCAP...02..007E, 2021arXiv210903246G}, as well as the small-scale measurements from the  Atacama Cosmology Telescope and South Pole Telescope \cite{2022arXiv220203515A}, together with the primordial abundance measurements of helium and deuterium. The CMB bound on DM mass is primarily driven by the measurement of $N_\mathrm{eff}$, and is stronger than the bounds derived from primordial abundance measurements from pristine Lyman-$\alpha$ systems, for most DM models. CMB-S4 will deliver a measurement of $\Neff$ with percent-level precision, $\sigma(\Neff)=0.03$, promising to put very stringent lower bounds on the mass of thermal-relic DM (or detect it), regardless of the details of its coupling to the Standard Model. Previous studies \cite{2022arXiv220203515A} found that the lower bounds on the mass from CMB-S4 will be able to exclude most of the mass range that could affect the process of BBN \cite{2020JCAP...01..004S,2022arXiv220203515A}. 

\section{Light (but massive) relics - {\it LiMR}s}

There have been many proposed extensions of the Standard Model that introduce the existence of light, weakly interacting particles. A general category is the one of light (but massive) relics (or LiMRs), which are particles that were in thermal contact
with the SM in the early universe and relativistic when they decoupled. The decoupling of these relics while relativistic gives these particles significant streaming motion, which sets a scale below which they cannot cluster, altering the large-scale structure (LSS) of our universe. 

One of the key signatures of these new light species is their contribution to the early universe energy density as radiation, which alters the expansion history of the universe and the epoch of radiation-matter equality, and additionally manifests in the CMB by modifying the Silk damping tail.  This signature of new dark radiation is commonly parametrized as $\Delta N_{\rm eff}$, and is quartically sensitive to the temperature of the light relic. In turn, measurements of the aforementioned scale-dependent suppression of large-scale structure are sensitive to both the mass and temperature of the LiMR \cite{DePorzio:2020wcz}, and thus our reach for new physics in this arena is greatly strengthened by joint analysis of CMB and LSS data.

A previous analysis of CMB temperature and polarization anisotropy measurements from the {\it Planck} satellite and full-shape LSS information from BOSS DR12 \cite{Xu:2021rwg} sets limits of $m_X <2.3~\mathrm{eV}~(2\sigma)$ for fermionic relics decoupling at TeV-scale temperatures. A notable application for this constraint is for light gravitinos in gauge-mediated SUSY scenarios, where the scale of SUSY breaking is bounded from above at $70~\mathrm{TeV}$. The considerable improvement of CMB anisotropy measurements expected with CMB-S4, with a notably sharpened sensitivity to $N_{\rm eff}$ contributions, would provide significant additional power towards detecting cosmic LiMRs (or the absence thereof). For the illustrative example of the gravitino, analysis of CMB-S4 and improved LSS data (such as from the forthcoming Dark Energy Spectroscopic Instrument - DESI) in tandem will yield unprecedented sensitivities: the ability to probe the entire parameter space at the $2\sigma$ level, and $3\sigma$ detection reach for all masses eV-scale and above. 

\section{Dark matter-baryon/electron interactions}
\label{sec:DM_baryon}

\begin{figure}[t!]
\begin{center}
\includegraphics[width=\columnwidth]{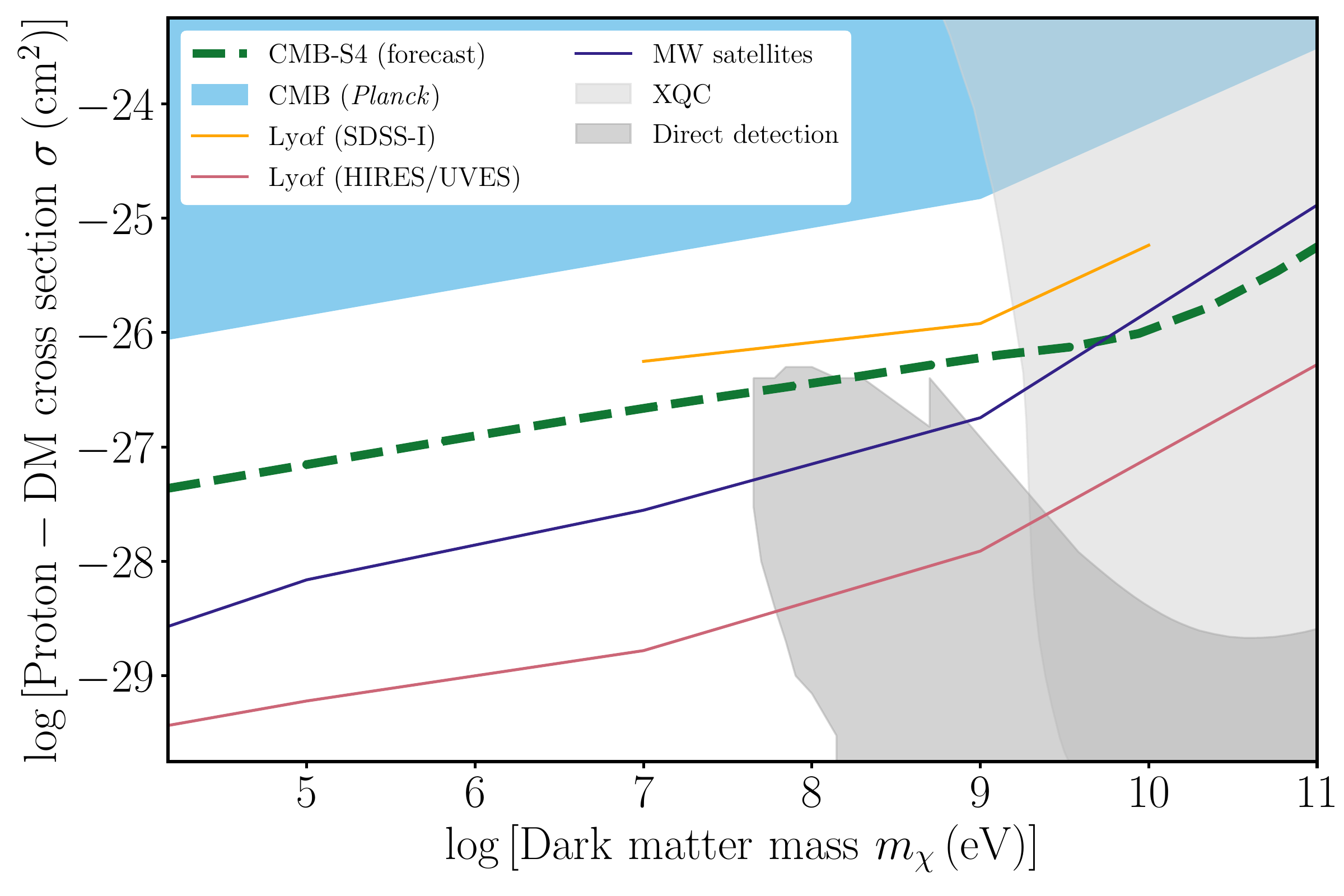}
\end{center}
\begin{center}
\caption{Dark matter (DM) -- proton cross section parameter space that can be probed with CMB-S4, for velocity-independent interactions. CMB-S4 is forecast \cite{Li:2018zdm} to improve existing \textit{Planck} CMB bounds \cite{PhysRevD.98.083510,Xu:2018efh,Dvorkin:2013cea} by over an order of magnitude. CMB-S4 is complementary to direct-detection searches \cite{PhysRevD.97.115047,2017EPJC...77..637A,2019PhRvD..99h2003A} and from the X-ray Quantum Calorimeter (XQC) \cite{2018JCAP...10..007M}, which have limited sensitivity to light (sub-GeV) and strongly-interacting DM. Large-scale structure limits come from the Lyman-alpha forest (Ly\(\alpha\)f) from SDSS-I \cite{Xu:2018efh} and from Keck-HIRES \& VLT-UVES spectra \cite{Rogers:2021byl}; and from the Milky Way (MW) satellite mass function \cite{2021ApJ...907L..46M,2021PhRvL.126i1101N,Buen-Abad:2021mvc}.\label{fig:bDM}}
\end{center}
\end{figure}

A generic and well-studied feature of dark matter models is some form of elastic interaction with Standard Model particles, in particular electrons~\cite{Nguyen:2021cnb,Buen-Abad:2021mvc} or nuclei~\cite{Chen:2002yh,Dvorkin:2013cea,Gluscevic:2017ywp,PhysRevD.98.083510,Xu:2018efh,Slatyer:2018aqg,Boddy:2018wzy,Rogers:2021byl}. These interactions are well constrained by terrestrial direct detection experiments, but cosmological data can serve as a powerful complement in several ways: firstly, to extend experimental sensitivity to smaller ($m_\chi \ll $ GeV) DM masses, where direct detection technologies are still developing; secondly, to increase our access to higher cross sections (such as would be expected for composite models) that would be lost to overburden and atmospheric scattering in underground experiments; and finally, in the case that a detection is made, to provide insight on how DM-SM interaction strength scales with energy and momenta transfer.

In the early universe, scattering with baryons or electrons may thermalize the dark matter distribution, washing out density fluctuations and impeding the growth of structure. The decreased clustering of matter imprints on the Cosmic Microwave Background as a suppression of temperature and polarization anisotropies at small scales, and the modified relative velocity between DM and the SM plasma further induces a shift in the scale of acoustic oscillations. As such, the increased precision of CMB-S4 temperature and polarization data, in addition to gravitational lensing at $\sim$arcmin angular scales \cite{Li:2018zdm}, will drastically improve the competitiveness of cosmological data in constraining dark matter physics, as well as access significant regions of parameter space that have been heretofore unprobed. For instance, for a velocity-independent cross section and 1 GeV dark matter mass, CMB-S4 will set a bound on the scattering with baryons of $\sigma < 6\times 10^{-27}$cm$^2$ at 95\% C.L. \cite{Li:2018zdm} (see Fig.~\ref{fig:bDM}).

\section{Dark matter freeze-in}

Dark matter could achieve its observed abundance via annihilation or decay of Standard Model particles in the early universe without thermalizing with the plasma. This is a generic mechanism known as DM ``freeze-in'' (see e.g. Refs.~\cite{Hall:2009bx,Bernal:2017kxu}). The parameter space for freeze-in represents the smallest relevant coupling the DM can have with the SM plasma that affects early Universe observables.
 If freeze-in DM couples to the Standard Model via a light mediator, most DM is produced at late times, and so in this sense freeze-in is insensitive to initial conditions. If the light mediator is the Standard Model photon or a kinetically-mixed dark photon (which is likely given strong constraints on other light mediators~\cite{Knapen:2017xzo,Green:2017ybv}), then DM becomes effectively charged. The freeze-in scenario is one of the few effective ways of making charged DM given the exclusion of the freeze-out scenario~\cite{McDermott:2010pa}. Freeze-in of effectively charged DM is a key target benchmark for many proposed and ongoing direct detection experiments~\cite{Essig:2011nj,Essig:2012yx,Essig:2015cda,Aguilar-Arevalo:2019wdi,Barak:2020fql,Amaral:2020ryn,Agnes:2018oej,Hochberg:2017wce,Geilhufe:2019ndy,Coskuner:2019odd,Knapen:2017ekk,Griffin:2018bjn,Griffin:2019mvc,Berlin:2019uco}. Additionally, since the DM is non-thermal in this scenario, freeze-in is one of the few allowed mechanisms for making sub-MeV DM from the SM plasma while still remaining consistent with bounds from BBN and $\Neff{}$ on thermal dark sectors (see the discussion in Section~\ref{sec:themalDM}). 

The portal responsible for DM production necessarily implies a DM-baryon interaction, creating effects that are similar to those described in Section~\ref{sec:DM_baryon} that are highly complementary to direct detection experiments. However, freeze-in requires a different analysis because scattering via a light mediator scales steeply with velocity as $v^{-4}$ and therefore the highly non-thermal phase space distribution must be taken into account to derive accurate limits~\cite{Dvorkin:2019zdi}. Previous studies showed that current constraints from CMB measurements from the {\it Planck} satellite set a freeze-in DM mass bound with $m_\chi \lesssim 18$ keV being excluded at the 95\% C.L., while CMB-S4 will be sensitive to masses up to $m_\chi \approx 29$~keV \cite{Dvorkin:2020xga}.

\section{Dark matter-neutrino interactions}

Dark matter interactions with Standard Model particles have received plenty of attention in the literature. One particularly intriguing scenario is the scattering of DM with Standard-Model neutrinos in the early universe \cite{2001PhLB..518....8B,2003astro.ph..9652B,2005A&A...438..419B,2010PhRvD..81d3507S,2006PhRvD..74d3517M,2015PhRvL.115g1304A,2018PhRvD..97d3513D,2021JCAP...03..066M}.
In this case, DM interactions lead to a suppression of the primordial density fluctuations through collisional damping \cite{2001PhLB..518....8B,2005A&A...438..419B}, which leaves noticeable signatures in the CMB anisotropy power spectra and the matter power spectrum, and ultimately affects the large-scale structure of the universe we observe today \cite{2014JCAP...05..011W,2018PhRvD..97g5039O,2021arXiv211004024H}. 

DM-neutrino interactions can affect CMB temperature and polarization power spectra by increasing the amplitude of the CMB acoustic peaks, and shifting them to higher multipoles $\ell$, with respect to the non-interaction scenario \cite{2014JCAP...05..011W,2021JCAP...03..066M,2021JCAP...10..017P}. Previous CMB analyses suggest that the constraints on cosmological parameters (in particular, DM–neutrino scattering cross section, and the sum of the neutrino masses), which rely primarily on measurements from \textit{Planck} at this time, could improve significantly with more accurate temperature and polarization data from the next-generation ground-based experiments. We expect the high-resolution measurements with CMB-S4 would increase the sensitivity to DM-neutrino interactions.

\section{Dark matter-dark radiation interaction}

In many theories (see e.g.~Refs.~\cite{Foot:2004pa,Kaplan:2009de,Kaplan:2011yj,CyrRacine:2012fz,Aarssen:2012fx,Cline:2013pca,Buen-Abad:2015ova,Lesgourgues:2015wza}), DM is part of an extended dark sector in which new interactions, such as a dark $U(1)$ force \cite{Ackerman:2008gi,Feng:2009mn,Agrawal:2016quu}, can couple the DM to relativistic degrees of freedom at early times. DM interacting with such dark radiation affects the CMB is two distinct ways. First, the presence of this extra dark radiation affects the expansion history of the universe, affects the Silk damping tail of the CMB (similarly to the impact of $N_{\rm eff}$), generates an early Integrated Sachs-Wolfe (ISW) effect, and modifies the redshift of matter-radiation equality. However, unlike the standard free-streaming neutrinos, this dark radiation is tightly-coupled (fluid-like) at early cosmological epochs, hence resulting in different phase and amplitude shift to the CMB acoustic peaks \cite{Bashinsky:2003tk,Cyr-Racine:2013jua,Follin:2015hya,Baumann:2015rya} for modes entering the causal horizon before the onset of dark radiation free-streaming.

Second, the large dark radiation pressure prohibits the growth of dark matter fluctuations on scales entering the horizon before kinematic decoupling of this DM-dark radiation fluid. On these characteristic length scales, the presence of dark acoustic oscillations (DAOs) and Silk-like damping decreases the depth of the gravitational potential near recombination, reducing the source term of CMB temperature fluctuations. In addition to these impacts on the primary CMB, the presence of DAOs and the small-scale damping of matter fluctuation will affect the lensing of the CMB photons as they travel towards us from the last scattering surface.

The DM-dark radiation models with the greatest deviations from the standard picture have already been constrained with data from the \textit{Planck} satellite~\cite{Cyr-Racine:2013fsa,Archidiacono:2019wdp}. In general, models in which all of the DM couples to dark radiation at the time when Fourier modes probed by the CMB enter the horizon are ruled out. This leaves scenarios in which only a small fraction of the DM couples to dark radiation as viable candidate to be probed by CMB-S4, which will strongly constrain any models that are currently still allowed given \textit{Planck} data.

\begin{figure}[t!]
\begin{center}
\includegraphics[width=0.75\textwidth]{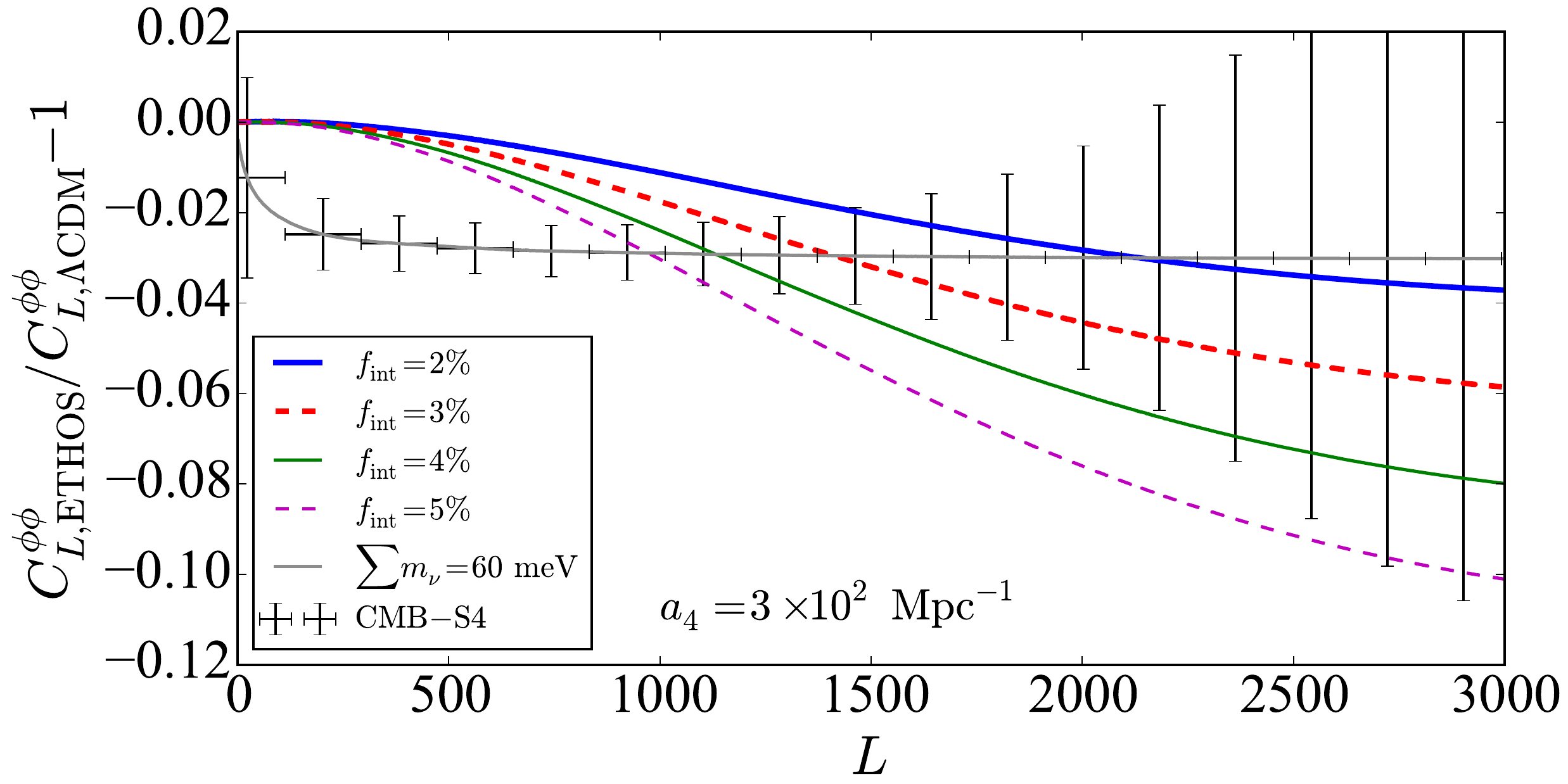}
\caption{ Fractional difference of the CMB lensing spectrum $C_L^{\phi\phi}$ between a standard $\Lambda$CDM model (with massless neutrinos) and four different interacting DM models (``ETHOS" models), each characterized by a different interacting fraction. For comparison, we also display a standard massive neutrino model with $\sum m_\nu =0.06$ eV.  The parameter $a_4 = 3\times 10^2$ Mpc$^{-1}$ characterizes the strength of the interaction between DM and dark radiation \cite{Cyr-Racine:2015ihg}.}\label{fig:Cls_phi_PIDM}
\end{center}
\end{figure}

As an illustration of the reach of CMB-S4, we show in Fig.~\ref{fig:Cls_phi_PIDM} the impact of different interacting DM models on the CMB lensing power spectrum. We can see here the fractional difference between interacting DM models and the standard $\Lambda$CDM case with the sum of neutrino masses $\sum m_\nu = 0$, for models where the fraction of interacting DM $f_\mathrm{int}$ is varied from $2\%$ to $5\%$, highlighting the fact that CMB-S4 will rule out models with fractions as low as $2\%$. Importantly, the interacting DM-dark radiation model predicts a different shape to the lensing power spectrum than that of massive neutrinos, highlighting that CMB-S4 will be able to distinguish between these two possibilities. 

\section{Axions and other ultra-light bosons}

Ultra-light axions (ULAs) are an attractive candidate for dark matter \cite{Matos:1992qx,Matos:1999et,Hu:2000ke,Suarez:2011yf,Hwang:2009js,Park:2012ru,Magana:2012xe} due to the fact that they behave like $\Lambda$CDM on large scales but also alleviate challenges to $\Lambda$CDM on small scales \cite{Hu:2000ke,Bullock:2017xww}. Furthermore, they can also behave as an early dark energy component to mitigate differences between high-redshift and local measurements of the Hubble constant $H_0$~\cite{Poulin:2018cxd}. 

The key feature of ultra-light axions dark matter is that they have de Broglie wavelengths that are macroscopic, which would generate cored galactic halo density profiles on scales of $\sim 0.7~(m_{\phi}/10^{-22}~{\rm eV})^{-1/2}~{\rm kpc}$, where $m_{\phi}$ is the mass of the axion. This feature has gone some way to resolving the cusp-core conundrum of Milky-Way dwarf spheroidal galaxies \cite{Lee:1995at,Hu:2000ke,Arbey:2003sj,Marsh:2013ywa,Marsh:2015wka,Gonzales-Morales:2016mkl,Bernal:2017oih,Deng:2018jjz,Broadhurst:2019fsl}. As discussed in Ref.~\cite{Marsh:2015wka}, a dynamical analysis of Fornax and Sculptor suggests that $0.3\times 10^{-22}~{\rm eV}\lesssim m_{\phi}\lesssim 10^{-22}~{\rm eV}$ can fit the profile of these dwarf spheroidal galaxies, however some tensions remain in solving both the ultra-faint dwarf satellites and the dwarf spheroidal population \cite{Safarzadeh:2020ku,Calabrese:2016po}. 

While baryonic effects have been suggested to mitigate challenges to $\Lambda$CDM at small scales, ULAs have been claimed to partially resolve these issues by suppressing the number of low-mass subhalos around Milky-Way scale halos and by the depression of satellite galaxy masses in halos relative to $\Lambda$CDM expectations \cite{Marsh:2013ywa,Marsh:2015xka,Marsh:2016vgj,Hui:2016ltb}. 

\begin{figure}[ht!]
\begin{center}
\includegraphics[width=\columnwidth]{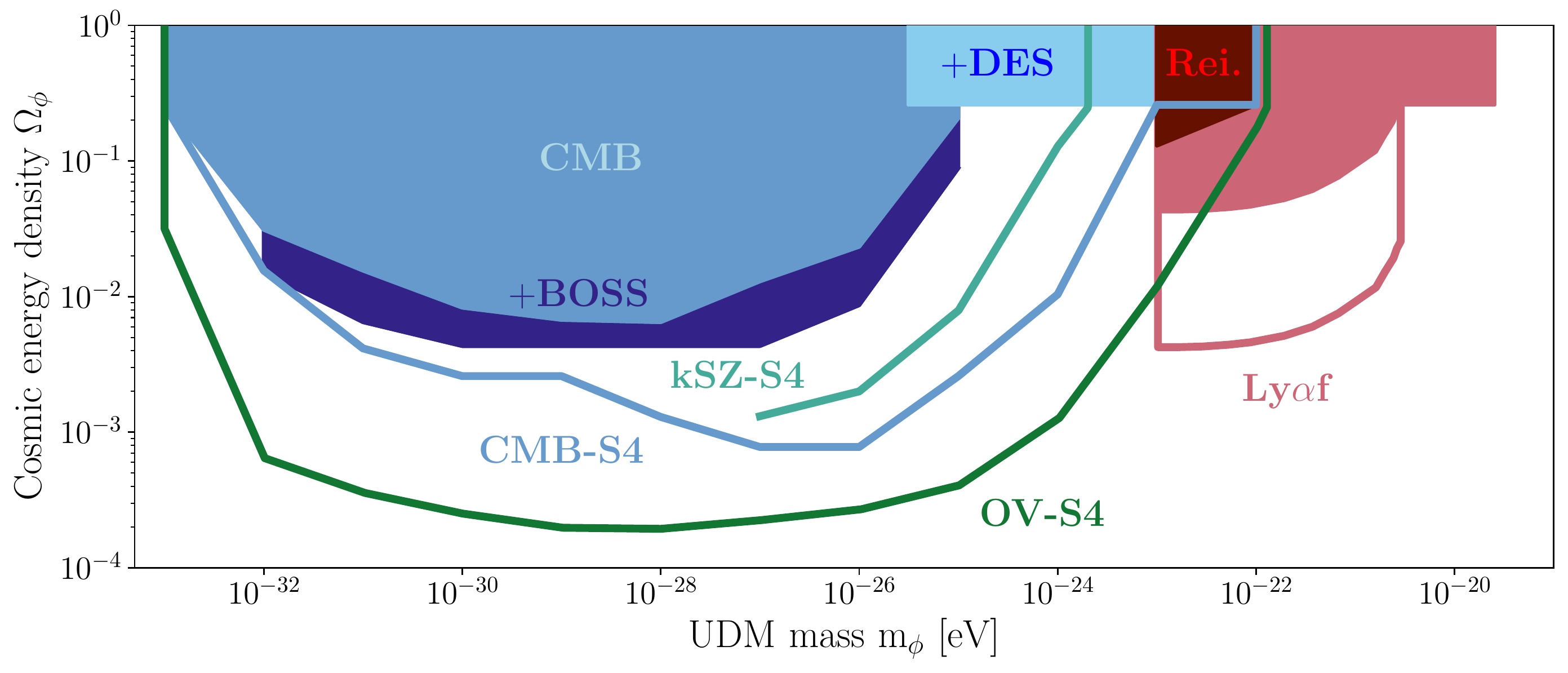}
\end{center}
\begin{center}
\caption{Ultra-light boson dark matter (UDM) parameter space that can be probed with CMB-S4, compared to other current (\textit{solid}) and projected (\textit{lines}) cosmological bounds. \textit{In turquoise}, UDM parameter space probed by kinetic Sunyaev-Zel'dovich measurements combining CMB-S4 data (including primary anisotropies) with large-scale structure information from the Dark Energy Spectroscopic Instrument (kSZ-S4) \cite{2021arXiv210913268F}; \textit{in blue}, forecast CMB-S4 UDM sensitivity using temperature, polarization and lensing data \cite{Hlozek:2016lzm}; \textit{in green}, forecast Ostriker-Vishniac effect bounds from CMB-S4 (OV-S4) \cite{2021arXiv210913268F}. Existing CMB bounds come from \textit{Planck} \cite{Hlozek:2014lca,Poulin:2018cxd}, while LSS bounds come from galaxy clustering combined with \textit{Planck} (+BOSS) \cite{2022JCAP...01..049L}, galaxy weak lensing combined with \textit{Planck} (+DES) \cite{2021arXiv211101199D}, the UV luminosity function and optical depth to reionization (Rei.) \cite{Bozek:2014uqa}, and the Lyman-alpha forest (Ly\(\alpha\)f) \cite{2021PhRvL.126g1302R,2020RogersPRD,Kobayashi:2017jcf,2017PhRvL.119c1302I,2017MNRAS.471.4606A}.\label{fig:ula}}
\end{center}
\end{figure}

Depending on their assumed mass, ULAs affect the CMB power spectrum on large scales through the ISW effect, and on small scales in the damping tail. The most sensitive probe of large-scale structure from measurements of the CMB is through the measurement of the lensing deflection on small scales. It is this regime that will see the largest improvement with future CMB experiments such as CMB-S4, as shown in Figure~\ref{fig:ula}. In addition, CMB photons may be Compton re-scattered by electrons along the line of sight. When this occurs in galaxy clusters, the resulting kinetic Sunyaev-Zel'dovich (kSZ) induces additional CMB temperature anisotropies that probe the real-time growth of structure, and can be detected via cross-correlation with galaxy surveys. ULAs will boost this signal (due to increased bias in a structure-suppressing cosmology), yielding a potential improvement in sensitivity \cite{2021arXiv210913268F}. Even better sensitivity is possible using the Ostriker-Vishniac (OV) CMB anisotropies, induced by gas inhomogeneities during the mildly non-linear reionization epoch.

In contrast to other DM candidates, the ULA field could be excited prior to thermalization of the cosmic plasma. In particular, if the $U(1)$ symmetry of ULAs is broken before the end of inflation, isocurvature fluctuations would be excited \cite{Axenides:1983hj,Turner:1983sj,Lyth:1991ub,Fox:2004kb,Hertzberg:2008wr,WMAP:2008lyn,Langlois:2010xc}, with an amplitude determined by 
the Hubble parameter $H_{I}$ during inflation and $m_{\phi}$ \cite{Visinelli:2009zm,Marsh:2013taa,Marsh:2014qoa,Visinelli:2014twa,Visinelli:2017imh}. If $H_{I}$ can be independently determined using CMB B-mode polarization anisotropies (e.g., by future experiments like CMB-S4 itself \cite{cmbs4}, Simons Observatory \cite{Galitzki:2018avl}, or LiteBIRD \cite{2018JLTP..193.1048S}) via the production of tensor modes \cite{Seljak:1996gy,Kamionkowski:1996ks}, a ULA energy density contribution of 
$\Omega_\phi \sim 0.01$ could be detected by CMB-S4 \cite{cmbs4,Hlozek:2017zzf}. Current constraints on isocurvature and tensor modes allow a fractional ULA  contribution of $\sim 10\%$ to the total DM budget \cite{Hlozek:2016lzm,Hlozek:2017zzf}. 

Alternatively, if the ULA $U(1)$ symmetry is broken after the end of inflation, a white-noise power spectrum of isocurvature would be produced. Depending on details of ULA production (such as thermal corrections to the ULA mass), CMB-S4 sensitivity to isocurvature fluctuations (when combined with CMB lensing, and independent probes of galaxy clustering and lensing shear) could test ULA masses as high as $m_{\phi}\simeq 10^{-17}~{\rm eV}$ \cite{Feix:2020txt}.

\begin{table}[ht!]
    \renewcommand{\arraystretch}{1.4}
    \footnotesize
    \centering
    \begin{tabular}{c|c}
    \hline 
        \textbf{Model} & \textbf{Limit}  \\
         \hline
         \hline
          DM annihilation & $m_{\chi}>30-50$ GeV 
        (thermal cross section, 
        20\% of energy deposition)\\
        \rowcolor[HTML]{EFEFEF} Thermal DM mass &  $m_{\chi}>10-15$ MeV (depending on DM models) \\
         LiMRs &   3$\sigma$ (non-)detection for $m_X>\mathcal{O}(1\hspace{1mm}{\rm eV})$, in combination with DESI\\
        \rowcolor[HTML]{EFEFEF} DM-baryon scattering &  $\sigma< 6\times 10^{-27}$cm$^2$ (for 1 GeV DM)\\
        Sub-MeV DM freeze-in  & $m_\chi>29$ keV\\
        \rowcolor[HTML]{EFEFEF} DM-DR interaction &$f_\mathrm{int} < 2\%$ \\
         ULA: $10^{-31}~\mathrm{eV} <m_\phi<10^{-24.5}$ & $\Omega_\phi <3\times 10^{-3}$\\
        \hline
    \end{tabular}
    \caption{Summary of sensitivity for different dark matter scenarios using CMB-S4.}
    \label{tab:summary}
\end{table}

\section{Conclusion}
 
CMB measurements provide a means to assess the impact of dark matter across the whole of cosmic history.  The sensitivity of the CMB measurements to the dark sector do not rely on assumptions about the local dark matter distribution and are insensitive to the details of astrophysical modeling that are necessary for some indirect searches.
We summarize the key results for the dark matter scenarios discussed in this paper in Table~\ref{tab:summary}.

CMB-S4 will probe deep into the damping tail of the CMB fluctuations, where much of the imprints from different dark matter scenarios deviates most strongly from the standard $\Lambda$CDM picture. Additionally, the improved measurements of the CMB lensing power spectrum will provide insights into the clustering of matter across a wide range of scales.  Measurements of matter clustering are valuable for deriving constraints on dark matter physics, since clustering is expected to be suppressed at various scales for many dark matter scenarios. 

CMB-S4 will dramatically improve the precision with which we measure CMB anisotropies. This leap in sensitivity will enable new insights into the nature of dark matter, complementary to what is expected from laboratory experiments and other astrophysical searches. 

\vspace{0.5cm}
\subsection*{Acknowledgments}
CD is partially supported by Department of Energy (DOE) grant \mbox{DE-SC0020223}. RH is a CIFAR Azrieli Global in the Gravity and the Extreme Universe program, and a Sloan Fellow. The Dunlap Institute is funded through an endowment established by the David Dunlap family and the University of Toronto. Canadian co-authors acknowledge support from the Natural Sciences and Engineering Research Council of Canada. Computations were performed on the SciNet supercomputer at the SciNet HPC Consortium. SciNet is funded by: the Canada Foundation for Innovation; the Government of Ontario; Ontario Research Fund - Research Excellence; and the University of Toronto. KB acknowledges support from the NSF Grant No.~PHY-2112884. FYCR was supported in part by the NSF grant No.~AST-2008696. GSF acknowledges support through the
Isaac Newton Studentship and the Helen Stone Scholarship at the University of Cambridge. VG and RA acknowledge the support from NASA through the Astrophysics Theory Program, Award Number 21-ATP21-0135. DJEM is supported by an STFC Ernest Rutherford Fellowship. JM is supported by the US~Department of Energy under Grant~\mbox{DE-SC0010129}. 

\bibliographystyle{unsrt2authabbrvpp}
\bibliography{main.bib}

\end{document}